# Wholeness as a Hierarchical Graph to Capture the Nature of Space

Bin Jiang

Faculty of Engineering and Sustainable Development, Division of Geomatics
University of Gävle, SE-801 76 Gävle, Sweden
Email: bin.jiang@hig.se

**Abstract**
According to Christopher Alexander's theory of centers, a whole comprises numerous, recursively defined centers for things or spaces surrounding us. Wholeness is a type of global structure or life-giving order emerging from the whole as a field of the centers. The wholeness is an essential part of any complex system and exists, to some degree or other, in spaces. This paper defines wholeness as a hierarchical graph, in which individual centers are represented as the nodes and their relationships as the directed links. The hierarchical graph gets its name from the inherent scaling hierarchy revealed by the head/tail breaks, which is a classification scheme and visualization tool for data with a heavy-tailed distribution. We suggest that (1) the degrees of wholeness for individual centers should be measured by PageRank (PR) scores based on the notion that high-degree-of-life centers are those to which many high-degree-of-life centers point, and (2) that the hierarchical levels, or the ht-index of the PR scores induced by the head/tail breaks can characterize the degree of wholeness for the whole: the higher the ht-index, the more life or wholeness in the whole. Three case studies applied to the Alhambra building complex and the street networks of Manhattan and Sweden illustrate that the defined wholeness captures fairly well human intuitions on the degree of life for the geographic spaces. We further suggest that the mathematical model of wholeness be an important model of geographic representation, because it is topological oriented that enables us to see the underlying scaling structure. The model can guide geodesign, which should be considered as the wholeness-extending transformations that are essentially like the unfolding processes of seeds or embryos, for creating beautiful built and natural environments or with a high degree of wholeness.

## 1. Introduction

It is commonly understood that science is mainly concerned with discovery, but only to a lesser extent, with creation. For example, physics, biology, ecology, and cosmology essentially deal with existing things in the physical and biological world and the universe, whereas architecture, music, and design are about creating new things. This polarization between science and the humanities, or between scientists and literary intellectuals, often referred to as the two cultures (Snow 1959), still persists, despite some synthesis and convergence (Brockman 1996). However, significant changes have happened. First, the emergence of fractal geometry (Mandelbrot 1989) created a new category of art for the sake of science (Mandelbrot 1989, Pertgen and Richter 1987). All those traditionally beautiful arts, such as Islamic arts and carpet weaving, medieval arts and crafts, and many other folk arts and architecture, found a home in science. Fractal geometry and chaos theory for nonlinear phenomena constitute part of a new kind of science called *complexity science*. The second change is large amounts of data, so called big data (Mayer-Schonberger and Cukier 2013), harvested from the Internet and, more recently, from social media such as Facebook and Twitter. This data has created all kinds of complex patterns, collectively known as visual complexity (Lima 2011). These two changes are closely interrelated. On the one hand, fractal geometry is often referred to as the geometry of nature, being able to create generative fractals that mimic nature, such as mountains, clouds, and trees. On the other hand, big data are able to capture the true picture of society and nature. In essence, nature and society are fractal, demonstrating the scaling pattern of far more small things than large ones. Both the generative fractals and visual complexity can consciously or unconsciously evoke a sense of beauty in the human psyche.

This kind of beauty evoked by fractals and visual complexity is objective, exists in the deep structure of things or spaces, and links to human feelings and emotions (Alexander 1993, 2002–2005, Salingaros 1995). The feeling is not idiosyncratic, but as a connection to human beings. It sounds odd that beauty is objective, because beauty is traditionally considered to be in the eye of the beholder. The beauty is an objective phenomenon, i.e., objectively structural, but we human beings do have subjective experience of it which may vary. While attempting to lay out the scientific foundation for the field of architecture, Alexander (2002–2005) realized that science, as presently conceived, based essentially on a positivist's mechanical world view, can hardly inform architecture because of a lack of shared notion of value. This is why most 20th-century architecture created all kinds of slick buildings, which continued into the 21st century in most parts of the world. Under the mechanical world view, feeling or value is not part of science. The theory of centers (Alexander 2002–2005) adopts some radical thinking, in which shared values and human feelings are part of science, particularly that of complexity science. In this theory of centers, wholeness is defined as a global structure or life-giving order that exists in things and that human beings can feel. What can be felt from the structure or order is a matter of fact rather than that of cognition, i.e., the deep structure that influences, but is structurally independent of, our own cognition. To characterize the structure or wholeness, Alexander (2002-2005) in his theory of centers distilled 15 structural properties to glue pieces together to create a whole (see Section 2 for details), and described the wholeness as a mathematical problem yet admitted in the meantime no mathematical model powerful enough to quantify the degrees of wholeness or beauty.

This paper develops a mathematical model of wholeness by defining it as a hierarchical graph, in which the nodes and links respectively represent individual centers and their relationships. The graph provides a powerful means for computing the degree of wholeness or life. First, the graph can be easily perceived as a whole of interconnected centers, enabling a recursive definition of wholeness or centers. Second, spaces with a living structure demonstrate a scaling hierarchy of far more low-degree-of-life centers than high-degree-of-life ones. The life or beauty of individual centers can be measured by PageRank (PR) scores (Page and Brin 1998), which are based on a recursive definition that high-degree-of-life centers are those to which many high-degree-of-life centers point. For the graph as a whole, its degree of life can be characterized by the ht-index derived from the PR scores; the higher the ht-index, the higher degree of life in the whole. The ht-index (Jiang and Yin 2014) was initially developed to measure the complexity of fractals or geographic features in particular, and it was actually induced by head/tail breaks as a classification scheme (Jiang 2013a), and a visualization tool (Jiang 2015a). Things of different sizes can be ranked in decreasing order and broken down around the average or mean into two unbalanced parts. Those above the mean, essentially a minority, constitute the head, and those below the mean, a majority, are the tail. This breaking process continues recursively for the head (or the large things) until the notion of far more small things than large ones is violated.

The contribution of this paper can be seen from several aspects. We illustrate the 15 structural properties using a generative fractal and an urban layout based on the head/tail breaks. We define wholeness as a hierarchical graph to capture the nature of space, with two suggested indices for measuring the degrees of life: PR scores for individual centers, and ht-index for a whole. The mathematical model of wholeness captures fairly well human intuitions on a living structure, as well as Alexander's initial definition of wholeness. Through the head/tail breaks, this paper helps bridge fractal geometry and the theory of centers towards a better understanding of geographic space in terms of both the underlying structure and dynamics. The mathematical model of wholeness can be an important model for geographic representation in support of geospatial analysis, since it goes beyond the current geometric and Gaussian paradigm towards topological and scaling thinking.

The remainder of this paper is structured as follows. Section 2 illustrates the 15 structural properties using the Koch snowflake and a French town layout. Section 3 defines the wholeness as a hierarchical graph and suggests how to quantitatively measure degrees of life for individual centers and the whole. Section 4 presents three case studies applied to an architectural plan and street networks of a city and country for measuring degrees of life or beauty in geographic spaces. Section 5 further discusses the mathematical model of wholeness related to beauty, creation/design, big data, and complexity science. Finally, Section 6 draws conclusions and points to future work.

**2. The 15 properties**
Following his classic work of the pattern language, and over the course of 30 years, Alexander (2002–2005) distilled 15 profound structural properties (Table 1) that can help generate all kinds of "good" patterns (e.g., Alexander et al. 1977, Thompson 1917) with so called living structure. The living structure comprises many, if not all, of these 15 fundamental properties, and therefore it possesses has a high degree of life, beauty, and wholeness. We can judge which one has a high degree of life by putting a pair of patterns or things side by side, such as the two snowflakes shown in Figure 1. This is the mirror-of-the-self test that uses human beings as a measuring instrument (Alexander 2002–2005, Wu 2015). In this connection, human judgment or feeling is not idiosyncratic, but reliable evidence because such a feeling is shared by a majority of people. Despite of the empirical evidence, the theory of center has received some harsh criticisms, e.g., the concept of life is accused of being subjective, and little experimental evidence to prove the theory (Alexander 2003). Alexander (2005) himself admitted that the 15 properties are somewhat elusive and hard to grasp. However, the 15 properties are found of great use to visual aesthetic research and design for a better built environment (Alexander 1993, Mehaffy and Salingaros 2015). This paper first illustrates the 15 properties using the Koch snowflake and a French town layout (Figure 2) as working examples.

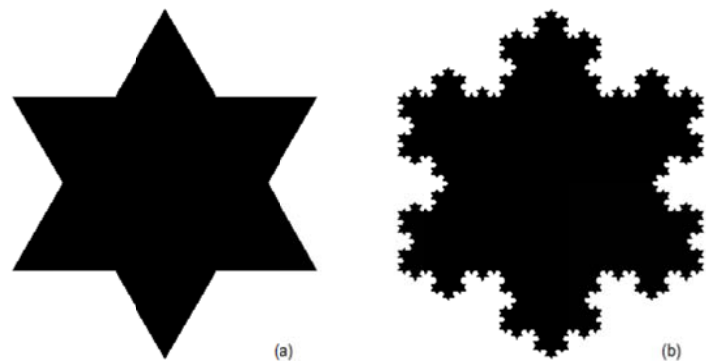

Figure 1: Koch snowflake with different degrees of life: Pattern b shows a higher degree of life (or complexity) than pattern a.
Note: Pattern (a) or Euclidean shapes in general appears to be cold and dry in terms of Mandelbrot (1982), while pattern (b) possesses a living structure according to Alexander (2002-2005) due to the presence of many of the 15 structural properties.

Table 1: The 15 structural properties

| Levels of scale | Good shape | Roughness |
| --- | --- | --- |
| Strong centers | Local symmetries | Echoes |
| Thick boundaries | Deep interlock and ambiguity | The void |
| Alternating repetition | Contrast | Simplicity and inner calm |
| Positive space | Gradients | Not separateness |

*Levels of scale*
As the building blocks of a whole, centers are defined at different levels of scale. For example, the snowflake has four scales: 1, 1/3, 1/9, and 1/27 (Figures 2a, 2b), while the axial map has five scales based on head/tail breaks (Figure 2f). In general, the levels of scale can be characterized by the ht-index (Jiang and Yin 2014), or the number of times that the scaling pattern of far more small things than large ones recurs.

*Strong centers*
A strong center is supported by other surrounding centers in a configuration as a whole. Centers are not separable in forming a coherent whole. The strongest center of the snowflake is in the middle of the pattern (Figure 2b), and it is supported by six, 18, and 54 other centers in a recursive manner. The strongest center of the axial map is the red line, and it is recursively supported by yellow, green, cyan, and blue lines (Figure 2f).

*Thick boundaries*
Centers are often differentiated by thick boundaries. For example, the different triangle sizes in the snowflake (Figure 2a) and the convex spaces of the urban layout have thick boundaries (Figure 2d). The five hierarchical levels of the axial map can be perceived as centers, represented by five different colors (Figure 2f), which apparently lack thick boundaries. In this regard, the different means used for the head/tail breaks process might be considered thin boundaries.

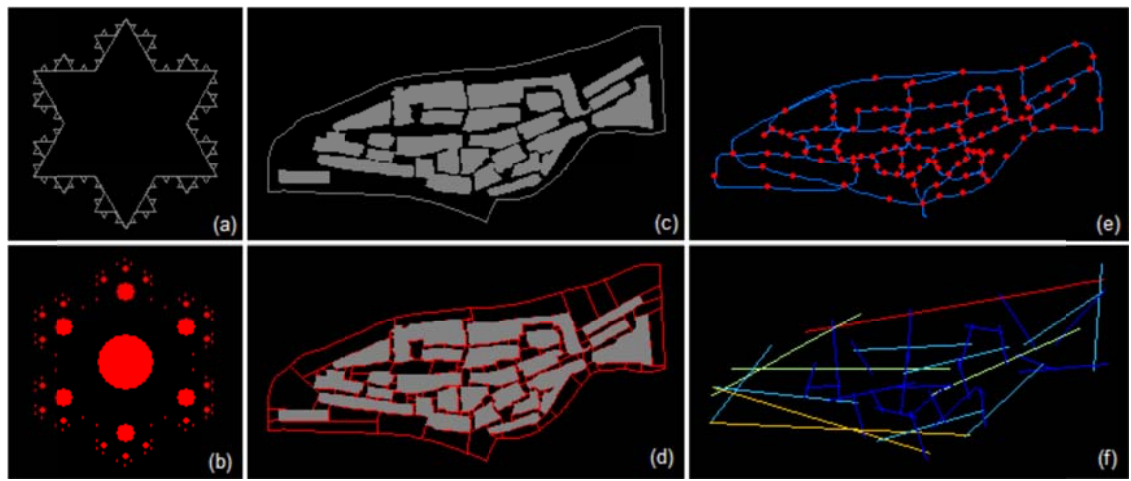

Figure 2: (Color online) Koch snowflake and the French town of Gassin: (a) the snowflake contains four levels of scale: 1, 1/3, 1/9, and 1/27 with thick boundaries; (b) the same snowflake without boundaries but with identified centers of different sizes; (c) the plan of the town of Gassin; (d) the space between building blocks is partitioned into many positive or convex spaces - the black polygons with red borders; (e) the positive spaces are perceived or represented as individual centers (red dots), and their adjacency (blue lines) forming a beady ring structure (Hillier and Hanson 1984); and (f) the related positive spaces are approximated into axial lines, which can be perceived as individual centers. The axial lines are in five hierarchical levels, perceived as five centers based on head/tail breaks, with red as the longest lines, blue as the shortest lines, and other colors for the lines between the longest and the shortest. It should be noted that the representation axial lines is better than that of convex spaces, because connectivity of axial lines rather than that of convex spaces demonstrate the scaling pattern of far more small things than large ones. This can be clearly seen in Panels e and f. The convex spaces have to be further aggregated into the axial lines in order to capture wholeness of the town plan.

*Alternating repetition*
Centers are strengthened if they are repeated by the property of alternating repetition. This property exists in the snowflake with the surrounding alternations of indents and outcrops of the edge, as well as in the axial map. The shortest lines (blue) bearing this property of alternating repetition are strengthened to form the lowest hierarchical level as a center, which supports other hierarchical levels or centers (Figure 2f). The notion of far more short lines than long ones recurs four times, also indicating an alternating repetition, or repeating the head/tail contrast statistically rather than strictly.

*Positive space*
The concept of positive space applies to both the figure and ground of a space. This is obvious in the town layout (Figure 2c), in which all the convex spaces are positive (Figure 2d), represented as individual centers forming a beady ring structure (Figure 2e). An axial line is an approximation of a set of adjacent positive spaces along a same direction; refer to Hillier and Hanson (1984) for more details.

*Good shape*
The concept of good shape is one of the most difficult properties to grasp. Alexander (2002–2005) suggested a recursive rule, in which parts of any good shape are always good shapes themselves. This sounds very much like self-similarity or alternating repetition. The snowflake is a good shape because it consists of many good triangular shapes (Figure 2a). The axial map is a good shape because it consists of many good shapes of axial lines (Figure 2f).

*Local symmetries*
Local symmetries refer to symmetries at individual levels of scale, rather than only at the global level. The snowflake shows both local and global symmetry (Figure 2a). The Alhambra plan (c.f., Section 3) is a very good example of local symmetries, so has a higher degree of life than that of the snowflake (Figure 2a) and axial map (Figure 2f).

*Deep interlock and ambiguity*
The figure and ground can be hardly differentiated while looking at the snowflake along its boundary (Figure 1b) because they interpenetrate each other, forming a deep interlock. This property of deep interlock and ambiguity is closely related to figure-ground reversal in Gestalt psychology (Rubin 1921). The same phenomenon appears in the town layout (Figure 2c and 2d), in which the building blocks and the pieces between them interpenetrate each other, creating ambiguity in visual perception.

*Contrast*
Contrast recurs between adjacent centers, thus strengthening the related centers. This kind of contrast appears between a big dot and its surrounding small dots (Figure 2b), and between the building blocks and the positive spaces (Figure 2d). There are many other different pairs of contrast, such as between the head and tail, red and blue, warm and cold colors, and a minority in the head and a majority in the tail.

*Gradients*
The centers gradually strengthen from the smallest to the largest scale, from the shortest to the longest line, from blue to red (Figure 2f), from the smallest to biggest dots (Figure 2b), and from the least-connected to the most-connected lines. This property of gradients can also be referred to as the scaling hierarchy ranging from the smallest to the largest.

*Roughness*
Roughness is what differentiates fractal geometry from Euclidean geometry and is synonymous to messiness or chaos. The border of the snowflake (Figure 2a) is rough, and an axial line is a rough representation of individual set of convex spaces (Figure 2f). Things with roughness may look messy or chaotic, but they possess a degree of order. It should be noted that both mathematical snowflake (Figure 2a) and real snowflake are rough, yet the former being strictly rough, while the latter statistically rough.

*Echoes*
The property of echoes can be compared to that of self-similarity in fractal geometry. In the snowflake, the triangle shape echoes again in different parts and in different sizes (Figure 2a). The scaling pattern of far more short lines than long ones (or the head-tail contrast) recurs or echoes four times (Figure 2f).

*Void*
*Void* is defined as an empty center at the largest scale, surrounded by many other smaller centers. Under this definition, the largest center in the snowflake is a void (Figure 2b), as is the longest axial line (Figure 2f). In general, the highest class (which may involve multiple elements) induced by the head/tail breaks constitutes a void.

*Simplicity and inner calm*
The degree of life of a center depends on its simplicity and inner calm, or the process of reducing the number of different centers. This process can be achieved through the head/tail breaks by grouping similar centers to one class. The six, 18, and 54 red dots can be clustered into three classes or centers (Figure 2b). This is the same for the axial map, which is classified into five hierarchical levels or centers (Figure 2f). Within each class, there is a sense of simplicity and inner calm.

*Not-separateness*
A center is not separable from its surrounding centers. This property has several other meanings. All scales, from the smallest to the largest, are essential and not separable in forming a scaling hierarchy. A whole connects to the nearby wholes to recursively form even larger wholes, toward the entire universe. A whole connects to human beings in their deep psyche, evoking a sense of beauty. That is why both the snowflake (Figure 2a) and axial map (Figure 2f) look beautiful.

The 15 properties bind all centers together into a whole to develop a high degree of wholeness. The more the properties, the higher the degree of life or the wholeness. We have shown in Figure 2 and through our elaboration that these properties are real and identifiable. To further quantify the degree of life, we define the wholeness as a hierarchical graph to measure the degree of life or wholeness.

## 3. Wholeness as a hierarchical graph

The idea of wholeness has been discussed in a variety of sciences such as physics, biology, neurophysiology, medicine, cosmology, and ecology (e.g., Bohm 1980), but no one prior to Alexander (2002-2005) has ever formulated and defined it in precise mathematical language. Following Alexander's definition of wholeness, some previous efforts have been made (e.g., Salingaros 1997) to quantify the degree of life of architecture. The proposed measure L does indicate approximately the degrees of life, but it lacks of the recursive property. We represent a whole as a graph, in which the nodes and links represent identified centers and their relationships within the whole (Figure 3). With the graph, we can compute the degrees of life for the individual centers and the whole. What is unique for our model is that it captures fairly well the recursive nature of wholeness as defined by Alexander. This section presents the two measures, the PR scores and ht-index, and argues why they can be a good proxy of degrees of life or beauty. In the next section, we further illustrate through case studies that a living structure demonstrates a scaling hierarchy of far more low-degree-of-life centers than high-degree-of-life centers; and the degree of the scaling hierarchy can be characterized by the ht-index: the higher the ht-index, the higher degree of life or wholeness.

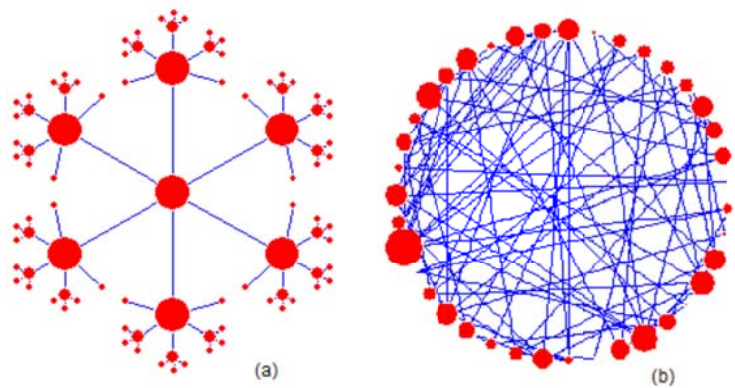

Figure 3: (Color online) Wholeness as a hierarchical graph of (a) the snowflake, and (b) the axial map
Note: The dot sizes indicate degree centrality in Figure 3a and betweenness centrality in Figure 3b.

## 3.1 Measuring the degrees of life using the PageRank scores for the centers

According to the theory of centers (Alexander 2002–2005), the degree of life of a center is strengthened by its surrounding centers. More importantly, the life (or intensity) has recursive or cumulative properties, implying that the degree of life of a center relies on those of all other centers in a whole. Given the recursion inherent in wholes or centers, the degree of life can be compared to social status of a person within a social network (Katz 1953, Bonacich 1987). To assess the social status of a person, we should not just ask how many people this person knows (which is the degree centrality), but also who are the people. In other words, it is not only popularity, but also social power or prestige that determines the person's status. This thought underlies Google's PR algorithm.

The definition of PR can be expressed as important pages to which many important pages point. This definition is recursive. To make an analogue, our importance depends on the importance of our friends, our friends of friends, and so on until virtually all people on the planet are included. It sounds very computationally intensive because it involves all the nodes of a graph. However, the computation is iterative until convergence is reached (see Langville and Meyer 2006 for more details). The PR model operates on a web graph, in which directed links indicate hotlinks from one page to another. Essentially, voting for incoming links determines the importance and relevance of individual pages, and therefore the PR scores. However, the voting is not one page one vote, but important pages have more votes. Formally, PR is defined as follows:

$$r(i) = \frac{1-d}{n} + d \sum_{j \in ON(i)} \frac{r(j)}{n_j} \qquad [1]$$

in which $n$ is the total number of nodes; $ON(i)$ is the outlink nodes (those nodes that point to node $i$); $r(i)$ and $r(j)$ are PR scores of nodes $i$ and $j$; $nj$ denotes the number of outlink nodes of node $j$; and $d$ is the damping factor, which is usually set to 0.85.

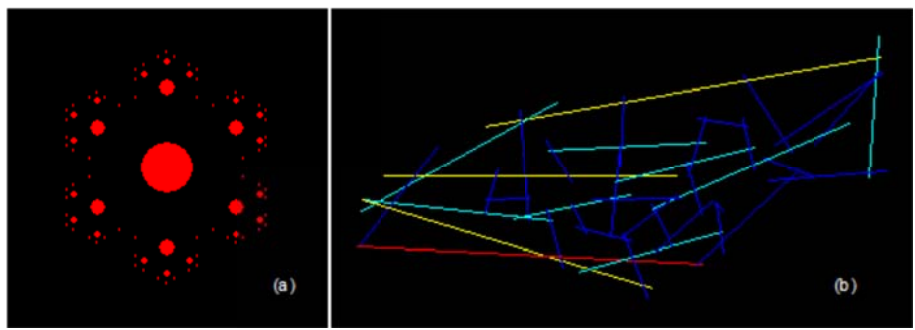

Figure 4: (Color online) Degrees of life of (a) the snowflake and (b) the axial map
Note: The degrees of life are visualized by dot sizes and the spectral colors, with red as the highest degree, blue as the lowest degree, and other colors as degrees between the highest and lowest.

The PR scores capture the spirit of wholeness, or the degree of life. We suggest a directed graph, in which surrounding centers point to a central one for computing the PR scores. The degrees of life in the snowflake's centers look the same as their sizes (Figure 4). However the degrees of life of axial lines differ from their length shown previously because PR is recursively defined. To this point, all centers are assigned degrees of life measured by PR scores. Based on degrees of life of centers, we can derive the ht-index as an indicator for the degree of life of the wholeness.

## 3.2 Measuring the degree of life using ht-index for the wholeness

Ht-index is a head/tail breaks induced index for measuring the complexity of fractals or geographic features in particular (Jiang and Yin 2014). It reveals the number of inherent scales in a set or pattern. The snowflake has four scales 1, 1/3, 1/9, and 1/27, so the ht-index is 4. For most patterns in the real world, the scales are not as clear as in the snowflake. We must conduct the head/tail breaks process to derive the inherent scales. In the axial map for example, the length of the axial lines exhibits a heavy-tailed distribution. We divided all 39 lines around the average length (first mean) into two parts.

Those lines longer than the mean constitute the head, and those that are shorter than the mean are the tail. This head/tail breaks process continues recursively for the head until it is no longer a minority (for example, < 40 percent). Each time a whole or head is broken into two parts, we must make sure that the head is a minority or, alternatively, the mean must be valid. In other words, a mean is invalid if the resulting head is not a minority. The ht-index is the number of valid means plus one. Formally, the ht-index (h) is defined as follows:

$$h = m(r) + 1 \qquad\qquad\qquad [2]$$

in which *m(r)* is the number of valid means during the head/tail breaks process for the PR scores.

In the same way as the axial-line length, we run the head/tail breaks process using the centers' PR scores to derive *h* as an indicator of the degree of life or the wholeness. The higher the ht-index, the higher degree of life or the wholeness. According to our calculation, the axial map has a higher degree of life (h=5) than that of the snowflake (h=4).

Why can the ht-index indicate the degree of life or the wholeness? The ht-index is a measure of the scaling pattern, or recurring times of far more small things than large ones. Figure 1b looks more complex, or involves more scales, than Figure 1a. This complexity, or the number of scales involved, is what the ht-index refers to. Our intuition also supports the notion that the higher the ht-index, the higher degree of life or the wholeness. For example, the embryo and city become more complex with more scales involved during the formation and development (Figure 5).

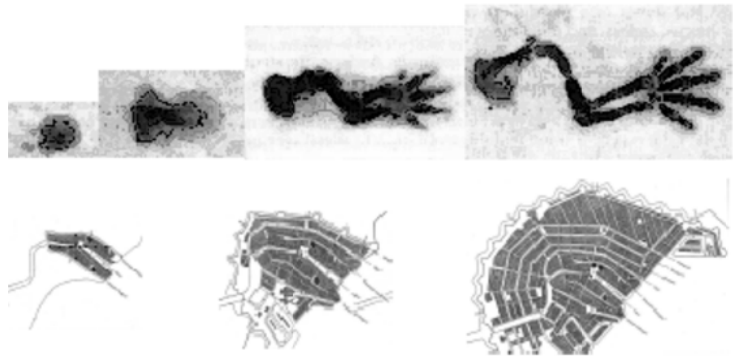

Figure 5: Intuitive examples of a higher ht-index meaning a higher degree of life
Note: A growing mouse foot in four days from day 12 to day 15, and the evolution of the city of Amsterdam from the 16$^{th}$ to the 18th century (Alexander 2002–2005)

**4. Case studies: Computing the degrees of life**
To further demonstrate the mathematical model of the wholeness, we applied it to three case studies in building, city, and country scales: the Alhambra building complex, the island of Manhattan (part of New York City), and the country of Sweden. We rely on the architectural plan of Alhambra, the axial map of Manhattan, and the street network of Sweden to compute their degrees of life. The first study used small data, with which we manually identified about 700 centers and more than 800 relationships between the centers for computation and analysis. The second and third studies involved large amounts of auto-generated axial lines and streets extracted from OpenStreetMap (OSM) databases. The two data sets were previously studied by Jiang (2013b) and Jiang et al. (2013) for cognitive mapping and map generalization.

**4.1 The plan of Alhambra**
The Alhambra is one of the most beautiful building complexes in the world and has a very high degree of life according to Alexander (2002–2005) and Salingaros (1997). We tried to compute the degree of life of the architectural plan and its numerous centers to see if our model of wholeness can capture the intuition or perception. Our study concentrated on its two-dimensional layout by ignoring the facades and internal structure in the vertical direction. The plan consists of three parts (left, right and middle), or nine subparts separated by red lines in Figure 6. We first manually drew all individual convex spaces from functional spaces such as rooms, courts, gardens, and halls, and then identified their relationships. The relationships are between two spaces that penetrate each other (such as with a connecting door), and are among the centers that belong to the same parts or subparts mentioned above. These convex spaces and their relationships constitute the nodes and links of a hierarchical graph to be used for computing the degrees of life.

The process of identifying the convex spaces and their relationships is time-consuming and tedious. The set of convex spaces must be the least number of fattest convex spaces; otherwise, the set would not be unique. The process starts with the first fattest convex space, the second fattest, and so on, until all spaces are covered (Hillier and Hanson 1984). As to their relationships, we must determine through visual inspection, i.e., which centers tend to support other ones? For example, peripheral centers support central ones. There is usually little ambiguity in terms of the relationships. We deliberately did not use any automatic process in order to make the relationships as accurate as possible for such small data. In this regard, this case study complements the following two cases involving big data, in which some automatic processes were adopted. Figure 6 shows the computed degrees of life, in which dots indicate the degree of life for the centers. The results are highly instructive. For example, the three centers with the highest degrees of life are rather obvious because of the recurring structural properties such as local symmetries, levels of scale, strong centers, thick boundaries, and positive space among the 15 properties. The ht-index of the wholeness is 6, derived from the degrees of life for all the centers, using equation [2].

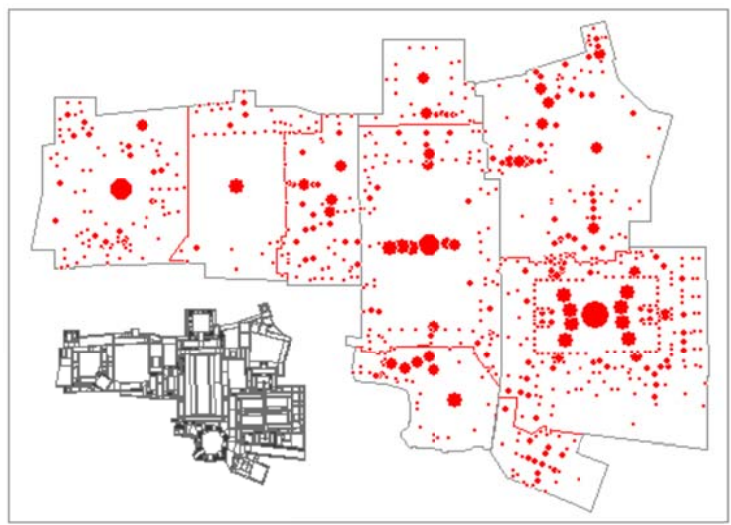

Figure 6: (Color online) The degrees of life computed for the plan of the Alhambra (inset)
Note: There are 725 convex spaces, or centers, and 880 relationships between the centers. The centers are divided into three parts (left, right, and middle) or subparts, indicated by red lines. The degrees of life for the individual centers are indicated by dot sizes, while the plan as a whole has a degree of life 6.

**4.2 The streets of Manhattan and Sweden**
The next two case studies move from the architectural scale to city and country scales. Manhattan as a city was chosen for the case study because its street network is easily perceived as a whole. We generated 1,800 axial lines for Manhattan and converted them into a hierarchical graph in terms of line-to-line intersection. This intersection relationship is based on the simple rule that short lines point to, or support, long ones. The same rule applies to the streets of Sweden. The streets are the complete set, including 166,479 streets for the entire country. The streets are a mix of named and natural streets (Jiang and Claramunt 2004, Jiang, Zhao and Yin 2008). In other words, individual street segments are merged according to the same names, and further merged together to form the natural streets. The street networks are then converted into dual graphs in which the nodes represent the individual streets, and the directed links indicate relationships from short streets (or axial lines) to long ones. Based on the directed graphs, we computed the degrees of life for the individual axial lines and streets, and they were visualized using the spectral colors, with blue as the lowest degree of life, red as the highest degree of life, and the other colors for degrees of life between the lowest and highest. It was surprising that the degrees of life demonstrated very striking power laws (Clauset et al. 2009), with two and three decades of power law fit, and a very high degree of goodness of fit (Table 2). The power law exponent around 2.0+ is consistent with the theoretic rules Salingaros (1998) suggested for scaling hierarchy.

Table 2: Power law statistics on PR scores for the Manhattan and Sweden networks
(Note: Alpha is the power law exponent, *P* an index for the goodness of fit, and *xmin* the minimum PR score, above which the power law is observed).

|  | Alpha | P | Xmin |
|---|---|---|---|
| Manhattan | 2.39 | 0.76 | 3.70E-04 |
| Sweden | 2.19 | 0.80 | 1.27E-05 |

Based on the degrees of life of individual centers, we further computed ht-indices of 5 and 7, respectively for Manhattan and Sweden each as a whole. The result appears to be consistent with our intuition that the Swedish street network is more alive than the Manhattan street network. This is because the former is far more heterogeneous and far larger than the latter. Manhattan as a city has rather simple grid-like layout, while the Alhambra as a building complex shows much more complex structure. Our calculation does support this intuition, i.e., the Alhambra has a higher degree of life than that of Manhattan, although the Alhambra has only 725 convex spaces, while Manhattan has 1,800 axial lines. This is in consistent with the definition of ht-index: the higher the ht-index, the more complex geographic features or space (Jiang and Yin 2014). From the case studies, we can foresee that modern buildings such as skyscrapers and urban structures such as shopping malls without much variation would have lower degrees of life or wholeness than traditional architecture.

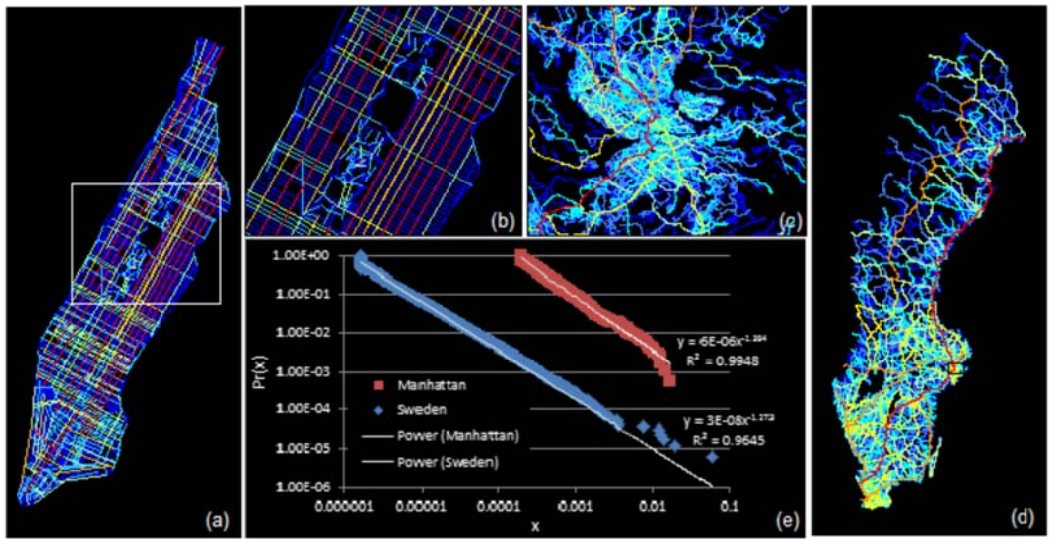

Figure 7: (Color online) The degrees of life shown through the underlying street structure of Manhattan (a) and Sweden (d)
Note: The two enlarged insets (b and c) present a better sense of far more low-degree-of-life centers than high-degree-of-life centers: Blue for the lowest degree of life, red for the highest degree of life, and other colors for degrees of life between the lowest and highest. The PR scores exhibit a very striking power law with two or three decades of power law fit and very high goodness of fit (p around 0.8) (e)

The three case studies demonstrate that the defined wholeness, or hierarchical graph, captures fairly well human intuitions on the degree of life. Various maps or patterns based on the head/tail breaks classification were found to have a higher degree of life than those using natural breaks through the mirror-of-the-self test (Wu 2015). It was also found that the kind of skills of appreciating the living structure can be improved through training. The reader might have noticed that the hierarchical graph goes beyond the exiting geographic representations (Peuquet 2002), since it is topological rather than geometric (Jiang and Claramunt 2004). The topological representation enables us to see the underlying scaling pattern, which constitutes a different way of thinking, the Paretian thinking, for geospatial analysis (Jiang 2015b). The model of wholeness would contribute fundamentally to geodesign by orienting or re-orienting it towards beautiful built and natural environments or with a high degree of wholeness. Geodesign, as currently conceived (e.g., Lee, Dias and Scholten 2014), mainly refers to a set of geospatial techniques and technologies for planning built and natural environments through encouraging wide-range human participation and engagement. However, there is a lack of standards in terms of what a good environment is. The model of wholeness provides a useful tool and indices for measuring the goodness. We therefore believe that geodesign should be considered as the wholeness-extending transformations, or something like the unfolding processes of seeds or embryos (Alexander 2002–2005) towards a high degree of wholeness (see a further discussion in Section 5). This idea of unfolding applies to map design as well, since maps are essentially fractal and possess the same kind of beauty (Jiang 2015c). This is in line with what we have discussed at the beginning of this paper that design should be part of complexity science. The mathematical model of wholeness also points to the fact that the wholeness or degree of life is mathematical and computational (Alexander 2002–2005), and it captures the nature of space, or geographic space in particular. The next section adds some further discussions on the mathematical model of wholeness and its implications in the era of big data.

## 5. Further discussions on the mathematical model of wholeness

Wholeness emerges from recursively defined centers, so it can be considered as an emergence of complex structures. To sense or appreciate the wholeness, we must develop both figural and analytical perception, or see things holistically and sequentially. However, a majority of people tend to see things analytically rather than figuratively (Alexander 2005). These two kinds of perception help us see a whole and its building-block centers, and perceive the degree of life or wholeness through the interacting and reinforcing centers. These perception processes are manifested in the mathematical model of wholeness. In other words, this model enables us to see things in their wholeness from their fragmented, yet interconnected, parts. The wholeness, or life, or beauty is something real, rather than a matter of opinion (Alexander 2002–2005). This kind of beauty exists in geographic space, arising from the underlying scaling hierarchy, or the notion of far more small geographic features than large ones (Jiang and Sui 2014). Large amounts of geographic information harvested from social media and the Internet enable us to illustrate striking scaling patterns (Jiang and Miao 2014, Jiang 2015a) and assess the goodness of geographic space.

The 15 properties are mainly considered as structural properties or the glue that holds space together, through which wholeness can be constructed. Recognition of the underlying structure is just one part of science - discovery. The other part is how to generate the kind of structure, or creation of the living structure, which is the central theme of the second book (Alexander 2002-2005). The 15 properties also can act as the glue for the creation, or the wholeness-extending transformations. For example, the process of generating the snowflake (Figure 1) is not additive but transformative. At each step, we do not just add smaller triangles, but transform the previous version as a whole, to give it more centeredness or wholeness by inducing more triangles (or centers in general) to intensify those that exist already. This generative process of the snowflake is the same as that of creating the life of the column in the step-by-step fashion elaborated by Alexander (2002-2005). In this regard, the mathematical model of wholeness is of use to guide the unfolding process because both the PR scores and ht-index provide good indicators for the degrees of life.

The mathematical model of wholeness is not limited to measuring the degree of life in geographic space. It can be applied to artifacts such as Baroque and Beaux arts, kaleidoscopes, and visual complexity generated from big data (Lima 2011). In spite of the popularity of the generative fractals and visual complexity, the question as to why they are beautiful has never been well-addressed. Through our model, we are able to not only explain why visual complexity and generative fractals are beautiful, but also measure and compare the degree of beauty. This kind of beauty exists in the deep structure, rather than in the surface coloring or appearance. This sense of beauty belongs to 90 percent of our self, or our feelings are all the same (Alexander 2002–2005). Importantly, the beauty has positive effects on human well-being. Taylor (2006) found that generative fractals, much like the natural scenes (Ulrich 1984), can help reduce physiological stress. Salingaros (2012) further argued that well-designed architecture and urban environments should have healing effects.

With the model of wholeness, the kind of beauty becomes computable and quantifiable. We mentioned earlier how the computed degree of life is consistent with human intuitions on living structures. One way to verify this is through eye-tracking experiments of human attention while watching a building plan such as the Alhambra (Yarbus 1967, Duchowski 2007). The captured fixation points from a group of people can be analyzed and compared to the degree of life computed using the model. This state-of-the-art methodology complements the mirror-of-the-self test, which only captures human intuitions on degree of life for a pair of patterns or things. The digital eye-tracking data can verify or compare our computed results on the degree of life. The beauty constitutes part of complexity science (Casti and Karlqvist 2003, Taylor 2003) and helps bridge science and arts in the big data era.

## 6. Conclusion

According to the theory of centers, all things and spaces surrounding us possess a certain degree of order or life, and those with a high degree of order are called living structures. This order fundamentally differs from what we are used to: regularity in terms of Euclidean geometry or normality in terms of Gaussian statistics. To put it more broadly, we are used to the 20th-century scientific worldview (mechanistic in essence), in which beauty is considered a matter of opinion, rather than that of fact. The living structure that exists in nature (e.g., Thompson 1917), as well as in what we build and make (e.g., Alexander et al. 1977) has many, if not all, of the 15 properties. This paper illustrated the 15 properties using two examples of space: the generative fractal snowflake and the French settlement layout. The illustration is well-supported by the head/tail breaks, a new classification scheme and visualization tool for data with a heavy-tailed distribution. We have shown the recurrences of the 15 properties in the living structures, making the 15 properties less elusive.

To quantify the living structure, this paper developed a model of wholeness based on Alexander's mathematical view of space. This model is a hierarchical graph in which numerous centers are represented by the nodes and their interactions are the directed links. Based on the initial definition of wholeness, particularly its recursive nature of centers, we suggested PR scores and ht-index as good proxies for the degrees of life because of their recursive nature. The three case studies presented some strong results. For example, the centers with the highest degrees of life in the Alhambra plan capture fairly well human intuitions on a living structure. More importantly, the degrees of life for both Manhattan's and Sweden's street networks demonstrate very striking power laws. These results are encouraging in terms of recognizing and appreciating the living structure. However, we are still far away from creating the kind of living structure known as the field of harmony-seeking computations (Alexander 2005). In this regard, we believe that the mathematical model of wholeness and related measures shed light on the wholeness-extending transformations. Our future work points in this direction.

**Acknowledgment**
The initial idea of this paper emerged while I was visiting the Tokyo Institute of Technology, supported by the JSPS Invitation Fellowship. I would like to thank Toshi Osaragi for hosting my visit, which I enjoyed very much. I also would like to thank Jou-Hsuan Wu and Ding Ma for helping with the case studies, Michael Mehaffy and the editor Brian Lees, and the three anonymous reviewers for their constructive comments.